\documentclass[11pt]{article}
\usepackage{array}
\usepackage{graphicx}
\usepackage{cite}
\usepackage{textpos}

\title{Three Easy Pieces \\
(in tribute to Roman Jackiw) \footnote{Contribution to {\it Roman Jackiw: 80$^{\rm th}$ Birthday Festschrift}, ed. A.~Niemi, T.~Tomboulis, K.~K.~Phua }}
\author{Frank Wilczek \\
\small\it Center for Theoretical Physics, MIT, Cambridge, MA 02139 USA; \\
\small\it T. D. Lee Institute and Wilczek Quantum Center, \\
\small\it Shanghai Jiao Tong University, Shanghai, China;\\
\small\it Arizona State University, Tempe, AZ, USA; \\
\small\it Stockholm University, Stockholm, Sweden}

\begin{document}

\maketitle

\begin{textblock*}{5cm}(11cm,-8.2cm)
\fbox{\footnotesize MIT-CTP/5167}
\end{textblock*}

\begin{abstract}
Roman Jackiw has made highly original and influential contributions to several areas of physics that have grown and blossomed, notably including the quantum physics of domain walls, magnetic monopoles, and fractional quantum numbers.   Here I offer three small pieces that take off from those themes.   I  discuss the emergence of topological surface structure in materials, the emergence of a shape-space magnetic monopole in a simple mechanical system, and the emergence of fractional angular momentum in an even simpler quantum mechanical (molecular) system.
\end{abstract}

\medskip

Roman Jackiw has had an uncanny knack for identifying ``curiosities'' that have grown into fertile, vibrant areas of physics research \cite{rj}.  His seminal contributions to the theory of anomalies, the interplay of topology with quantum theory, and fractional quantum numbers are a rich legacy which has become central to both fundamental physics and modern quantum engineering.   It has influenced my own work both directly and indirectly.   The direct influence should be obvious.  Among others, my work with Goldstone on fractional quantum numbers started as an attempt to understand, and then generalize, the remarkable discoveries of Jackiw and Rebbi \cite{jr} and of Su, Schrieffer, and Heeger \cite{ssh}; anyons and braiding arose from coming to terms with fractional angular momentum; axions arose as the possible solution to a big conceptual problem raised by the topology of gauge fields in QCD.   The indirect influence is perhaps less visible but no less profound.  I became aware of Dirac's recommendation to ``play with equations'' in my student days, but it came alive for me only a bit later, in large part because I saw how well it had worked for Roman.  That helped me feel liberated to indulge in such play.   

Here I will present three brief case studies in playing with equations, which have close ties to Roman's work.  Roman, I hope this little tribute brightens your day.   

\section{Domain Walls and Boundary Surfaces}

In this part I want to make a few simple but important observations which highlight the close connection of Jackiw and Rebbi's pioneering work to earlier problems and later developments, and may suggest other significant directions.   To make the discussion self-contained for both condensed matter and high energy physicists, I will start from scratch.

Consider a one-dimensional chain of identical molecules with spacing $a$.  We want to model the situation where there is one conduction electron per molecule.  Let us adopt the drastic simplifying assumption that the molecules simply provide a weak periodic background potential and ignore interactions among the electrons, aside from their mutual influence through fermionic quantum statistics.    To begin, we will also assume that the potential is spin-independent and invariant under spatial reflection.

According to standard band theory we should expect to have a metal, since the electrons exhaust only half the available states.  Indeed, the carrying capacity of the lowest band, allowing for the spin degree of freedom, is twice its extent in (quasi-)momentum space, and so the actual density corresponds to half filling:
\begin{equation}
N/L ~=~ 1/a ~=~  \frac{1}{2} \cdot 2 \cdot  \frac{\frac{2\pi}{a}}{2\pi} 
\end{equation}

%\begin{center}\label{peirelsMomentum}
%\includegraphics[scale=0.5]{peirelsMomentum.png}  
%\end{center}
%Figure \ref{peirelsMomentum}: a. Electrons at half filling in momentum space, for a one-dimensional lattice.  For concreteness, the drawing assumes that the lattice makes a weak perturbation to free electron behavior.   The Fermi sea is shaded.   Low energy excitations involve changes in occupancy near the Fermi points $k_F$, allowing us to linearize their dispersion relation.  b. The linearized spectrum can be mapped faithfully onto a light-cone.  c.  A deformation in the lattice structure which brings in momenta at $\pi/a$ can connect modes at $k_F + \xi$ and $-k_F + \xi$ -- right-moving particle and left-moving hole -- and open a gap in the dispersion relation.

%\begin{figure}[h!]
%\begin{center}
%\includegraphics[scale=0.5]{peirelsMomentum.png}
%\caption{a. Electrons at half filling in momentum space, for a one-dimensional lattice.  For concreteness, the drawing assumes that the lattice makes a weak perturbation to free %electron behavior.   The Fermi sea is shaded.   Low energy excitations involve changes in occupancy near the Fermi points $k_F$, allowing us to linearize their dispersion relation.  b. %The linearized spectrum can be mapped faithfully onto a light-cone.  c.  A deformation in the lattice structure which brings in momenta at $\pi/a$ can connect modes at $k_F + \xi$ and %$-k_F + \xi$ -- right-moving particle and left-moving hole -- and open a gap in the dispersion relation.}
%\label{peirelsMomentum}
%\end{center}

%\end{figure}

\bigskip

In the electrons' ground state, they occupy all the momentum states for $-k_F \leq k \leq k_F$ with
$k_F = \pi / 2a$.    The low-energy excitations around this ground state involve changes in the occupation of modes near the boundaries of that region.  There we can approximate the change in their energy with momentum -- that is, the deviation in momentum from $\pm k_F$ -- as a linear relationship, bringing in the Fermi velocity
\begin{equation}
v_F ~=~ \frac{dE}{dk}(k_F)
\end{equation}
Thus we derive, from the two points of the Fermi surface, left- and right-moving excitations obeying
\begin{eqnarray}
(\partial_t - \partial_x ) \psi_R ~&=&~ 0 \nonumber \\
(\partial_t + \partial_x ) \psi_L ~&=&~ 0 
\end{eqnarray}
where we adopt the unit of velocity $v_F \rightarrow 1$.  This can be summarized in relativistic form, as a Dirac equation
\begin{equation}
i \gamma \cdot \partial \, \psi ~=~ 0 
\end{equation}
with
\begin{equation}
\psi ~\equiv~ \left(\begin{array}{c}\psi_L \\\psi_R\end{array}\right) \, ; \ \gamma_0 ~\equiv~ \sigma_2 \, ; \ 
\gamma_1 ~\equiv~ i \sigma_1 
\end{equation}
This empowers us to visualize a Dirac cone.  

The modes from the two sides of the cone are separated, as we saw earlier, by $\Delta k = \frac{\pi}{a}$.   Insofar as that wavenumber is not represented in the background potential, those modes remain independent.  On the other hand, were the potential to contain that wave number it could connect the left- and right-movers, and open a gap in the Dirac cone.  This mixing pushes down the energies of occupied states, so it is a favorable effect, energetically.    We should inquire whether it can be triggered dynamically, and occur spontaneously.   

%\begin{center}\label{deformedLattice}
%\includegraphics[scale=0.35]{deformedLattice.png}\\
%${}$  \\ 
%\end{center}
%Figure \ref{deformedLattice}.  a. Vibration of a periodic lattice at $\pi/a$, corresponding to an ``optical'' phonon.  b. and c.: Condensation in the optical phonon mode leads to a deformation of the lattice, doubling its periodicity to $2a$.  The condensation can occur in two different ways, as shown.   d. A simple slip in the deformation process, indicated by the arrow, produces a configuration which looks like the ordering of b. to the right and like c. to the left.  Such a configuration can relax into a domain wall, which is stable for topological reasons.

%\begin{figure}[h!]
%\begin{center}
%\includegraphics[scale=0.30]{deformedLattice.png}
%\caption{a. Vibration of a periodic lattice at $\pi/a$, corresponding to an ``optical'' phonon.  b. and c.: Condensation in the optical phonon mode leads to a deformation of the lattice, %doubling its periodicity to $2a$.  The condensation can occur in two different ways, as shown.   d. A simple slip in the deformation process, indicated by the arrow, produces a %configuration which looks like the ordering of b. to the right and like c. to the left.  Such a configuration can relax into a domain wall, which is stable for topological reasons.}
%\label{deformedLattice}
%\end{center}
%\end{figure}

\bigskip

To generate the required potential, we need to let the molecules on neighboring lattice sites displace in alternate directions.    Displacements of the molecules correspond to phonons, and in that language we are asking whether the ``optical'' phonons at $k = \frac{\pi}{a}$ condense.  
If we denote the amplitude of the displacement (phonon)  $k = \frac{\pi}{a}$ field by $\phi (x, t)$, then we have a coupling
\begin{equation}\label{electronLagrangian}
{\cal L } ~=~ \bar \psi (i \gamma \cdot \partial - g \phi ) \psi 
\end{equation}
so that condensation in $\phi$ opens a gap and gives us a fermion mass $m = g\langle \phi \rangle$.   

This entire discussion can be phrased in the language of relativistic quantum field theory.  In that language, we are studying how mass generation through coupling to a scalar condensate back-reacts on the energetics of a scalar field which triggers it.   In particle physics this effect is called the Coleman-Weinberg mechanism \cite{cw}.  It plays an important role in the theory of the standard model, and in many speculations that go beyond the standard model.  

We can exploit that mapping to bring in the machinery of Feynman graphs, easing our computational challenge.   Our energy gain corresponds to changes in the one fermion loop vacuum amplitude brought in by changes in the fermion mass.    

%\begin{center}\label{feynmanEnergy}
%\includegraphics[scale=0.5]{feynmanEnergy.png}
%\end{center}
%Figure \ref{feynmanEnergy}: This one-loop Feynman graph contributes to the effective potential for the optical phonon mode.  It captures the favorable energetics of opening a gap (or, in relativistic language, mass generation).  

%\begin{figure}[h!]
%\begin{center}
%\includegraphics[scale=0.5]{feynmanEnergy.png}
%\caption{This one-loop Feynman graph contributes to the effective potential for the optical phonon mode.  It captures the favorable energetics of opening a gap (or, in relativistic %language, mass generation).}
%\label{feynmanEnergy}
%\end{center}
%\end{figure}

\bigskip

To assess the possible instability, we can focus on infinitesimal displacements.  The most singular part takes the form
\begin{eqnarray}\label{displacementEnergy}
{\cal E} ~&\sim&~ \int\limits^\Lambda_0 \, \frac{d^2 k}{(2 \pi)^2} \frac{ {\rm Tr} ( \gamma \cdot k )^2}{(k^2 + m^2)^2} \nonumber \\ 
~&\propto&~  (g \phi)^2 \, \ln \, (\Lambda / g\phi ) 
\end{eqnarray}
Note that it is appropriate to supply an ultraviolet cutoff, since our simple ``relativistic'' description of the electron modes is only valid for low-energy excitations, involving small deviations from the Fermi points.   

Of course, displacement of the molecules also involves ordinary, Hooke's law elastic energy.   But that energy, being proportional to $\phi^2$, is dominated by Eqn.\,(\ref{displacementEnergy}) for small $\phi$.   Thus the lattice will always deform.  This is Peirels' instability \cite{peirels}.  It is strikingly reminiscent of the Cooper instability which drives superconductivity.  

The most favorable displacement amplitude $\langle \phi \rangle \equiv \pm v$ has two possible values.  They correspond to two distinct homogeneous ground states.  They differ from one another by displacement through $a$ -- that is, by the transformation which used to be, but is no longer, a symmetry.  In these two configurations our electrons acquire, according to Eqn.\,(\ref{electronLagrangian}), effective mass $\pm gv$.   The negative value looks a bit strange, at first sight, but the sign of the mass can be absorbed, formally, into a redefinition of $\psi$.  Indeed, the transformation $\psi^\prime = \gamma_0 \gamma_1 \psi$ changes the sign of the mass term, while leaving the kinetic term invariant.   

A simple defect allows one to interpolate between the two distinct homogeneous states.  
%(See Figure \ref{deformedLattice}b.)  
%% EDIT: Should be \ref{deformedLattice} - an "m" is missing, so the reference is not compiling.
If we let the simple defect configuration relax, energetically, it will settle down to a minimum energy configuration -- a domain wall -- that is stable for topological reasons.   
In this configuration the effective electron mass $m(x)$ interpolates between $\, -gv$ (say) on the far left and $gv$ on the far right.   (The {\it relative \/} sign, of course, cannot be undone by re-definition.)  The effective mass will vanish somewhere in between, and we might anticipate that something interesting could happen as a result.    In fact. as discovered by Jackiw and Rebbi, there is a remarkably robust and characteristic consequence: the existence of zero energy (midgap) states, centered on the wall.   

Fixing the center of the wall to be at $x=0$, we have an effective mass $m(x)$ with $m(-x) = - m(x)$.  The electron equation for zero energy reads
\begin{eqnarray}
0 ~&=&~ (- i\gamma_1 \partial_x  + m(x) )  \psi (x) \nonumber \\
~&=&~(\sigma_1 \partial_x  + m(x) )  \psi (x)
\end{eqnarray}
Projecting this on spinor structures with $\sigma_1 s_{\pm} = \pm s_{\pm}$, we find that for $\psi(x) \equiv \psi_+ (x) s_+ \, + \, \psi_- (x) s_- $
\begin{equation}
(\partial_x \pm m(x) ) \psi_\pm (x) ~=~ 0
\end{equation}
with solutions
\begin{eqnarray}
\psi^0_+ (x) ~&=&~ e^{- \int\limits_0^x \, du \ m(u) } \label{goodSolution} \\
\psi^0_-(x) ~&=&~ e^{ \int\limits_0^x \,  du \ m(u) }
\end{eqnarray}
The first of these, i.e. Eqn.\,(\ref{goodSolution}) shrinks exponentially to zero in both directions away from the wall, and defines a normalizable state.  It is the zero energy, midgap state advertised just above.  

The existence of this zero energy mode has several interesting consequences.  It should be plausible, given the symmetry of the problem, that the zero energy mode draws half its strength from below and gap half from above.    Thus if we fail to occupy the zero mode, we have a deficit of half a unit of electron charge, relative to the ground state; while if we do occupy the zero mode, we have an excess of half a unit of electric charge.   One can also show this formally, either directly from the definition of the charge operator, or by calculating the flow of charge as one builds up the domain wall (see below).  Thus the domain wall induces a special kind of vacuum polarization, where a fractional charge $\pm \frac{1}{2}$ accumulates \cite{jr}.    

When we take into account the spin degree of freedom, the accounting takes a different form. Now there are two zero modes: one for spin up, one for spin down.   By occupying, or not, each of the two zero modes, we have:
\begin{itemize}\label{funnySpinCharge}
\item {\it zero occupancy}: charge $-e$, spin 0
\item {\it single occupancy}: charge zero, spin $\frac{1}{2}$ doublet
\item {\it double occupancy}: charge $+e$, spin 0
\end{itemize}
Though the charge spectrum is normal, the relation between charge and spin is unusual.   We can express the general situation through the equation
\begin{equation}\label{chargeSpinTopologyRelation}
(-1)^{2S} (-1)^{Q/e} ~=~ (-1)^W
\end{equation}
where $W$ is a topological quantity, indicating the number of domain walls.   Since the quantities that appear in this equation are intrinsically discrete the relationship it expresses will be exact, unless our approximations have been very bad.   Moreover, it is the only non-trivial relation among $2S, Q/e$, and $W$ consistent with the property that two domain walls can annihilate into states with normal quantum numbers.  We expect this property to be valid, because the lattice of molecules with two minimal defects differs from a correctly ordered (i.e., ground state) lattice only within a bounded region, and is topologically trivial.

The logic of the Peirels instability is not restricted to half filling \cite{ssh}.  One can, for example, consider conduction bands that fill $1/k$ of the available states, and trigger an instability toward charge density waves at $2\pi/k$, with $k$ an integer.   Then we will have domain walls that can annihilate in $k$-tuples.   If such annihilation is accompanied by emission or absorption of $l$ electrons, we can deduce a generalization of Eqn.\,(\ref{chargeSpinTopologyRelation}), in the form
\begin{equation}
e^{i2\pi (Q/e  - lW/k)} ~=~ 1
\end{equation}
(Here, for simplicity, I have not kept track of the spin quantum number, which need not be conserved separately.)  Equivalently, we can write
\begin{equation}\label{chargeWallRelation}
Q/e ~=~ l W  / k \, + \, {\rm integer}
\end{equation}
Earlier we had $k=2, l=1$, and half-integer charge.  In that case, we could understand the fractional charge based on the existence of a zero energy solution with equal particle-hole character.  But in more general situations, say for example $k=3, l=1$, where we have third-integer charge, it cannot be understood in that way.

An appropriate, minimal model for these more general situations allows the field $\phi$ in Eqn.\,(\ref{electronLagrangian}) to become complex.   The overall phase of $\phi$ adjusted by re-defining $\psi$, according to $\psi^\prime = e^{i\lambda} \psi$, but relative phases in $\phi(x, t)$ are physically meaningful.   In particular, we can have situations where there are domain walls which interpolate between values
\begin{eqnarray}\label{wallAsymptotics}
\phi(x) \, &\rightarrow& \, \ \ v \, ;  \ \ \ \ \ \ \  x \rightarrow - \infty \nonumber \\
\phi(x) \, &\rightarrow&  \, e^{2\pi i/ k} \, v \, ;  \ \ x \rightarrow + \infty
\end{eqnarray}
Such domain walls can annihilate in $k$-tuples.   We expect, based on the preceding discussion, that fractional charge may accumulate on such walls.  

An efficient way to calculate the charge is first to imagine building up configurations with the wall asymptotics of Eqn.\,(\ref{wallAsymptotics}) gradually, starting from trivial asymptotics \cite{gw}.    As long as the magnitude of the local gap exceed the field gradients, i.e.
\begin{equation}\label{softWall}
\frac{ | \partial \phi | }{ g |\phi |} \, << 1
\end{equation} 
there will be no particle production, and we can calculate the current flow (to lowest order in gradients) by means of a simple vacuum polarization graph, similar in spirit to how one calculates the correction to the phonon energy, but now with insertion of the electron number current in place of a phonon field. An elegantly simple result emerges, in the form 
\begin{equation}\label{gwCurrent}
\langle j^\mu \rangle \, = \, \frac{1}{2\pi} \, \epsilon^{\mu \nu} \, \partial_\nu \, {\rm Arg } \, \phi  
\end{equation}
and for the integrated charge
\begin{eqnarray}\label{gwRelation}
Q/e \, &=& \, \int\limits^\infty_{- \infty} dx j^0 \nonumber \\ 
\, &=& \,  \frac{1}{2\pi} \, \int\limits^\infty_{- \infty} dx \, \partial_x {\rm Arg} \, \phi \nonumber \\
\, &=& \, \frac{1}{2\pi} \, \bigl( {\rm Arg} \, \phi (\infty ) \,  - \, {\rm Arg } \, (- \infty) \bigr)
\end{eqnarray}

Now the realistic, minimum energy wall configuration may not satisfy Eqn.\,(\ref{softWall}) everywhere, though of course it does so asymptotically (where $\partial \phi \rightarrow 0$).   So we must imagine a second step, where we build up the steeper gradients.  During that process electrons can be radiated to, or absorbed from, infinity.  But since the electrons are normal out there, any such radiation will change the electron number by an integer.   Also, from a complementary perspective, the angle function ${\rm Arg} \phi$ becomes ill-defined at $\phi = 0$, and we need to allow for extra $2\pi$ jumps, as we integrate Eqn.\,(\ref{gwCurrent}) through such points.  For both these reasons, we should generally interpret Eqn.\,(\ref{gwRelation}) as a relation modulo integers, i.e. as a formula for the fractional part of the charge.  As such, it precisely embodies Eqn.\,(\ref{chargeWallRelation}).  

We can ``explain'' the existence of the zero energy solution we found earlier, based on these more global considerations, as follows.  The zero energy solution occurred in the model with $g$ and $\phi$ real.   Within that framework we cannot achieve the non-trivial domain wall asymptotics, where $\langle \phi \rangle$ change sign, without encountering a zero of $\phi$, where the requirement of Eqn.\,(\ref{softWall}) cannot be met.   We can get around that difficulty by adding a small infinitesimal imaginary piece to $\phi$, and then removing it at the end.   Since there is a gap, the limit is harmless.   But depending on the sign of the added piece, we will get $Q/e = \pm \frac{1}{2}$.   So there must be degenerate states with those quantum numbers, and therefore a zero energy mode of the electron field, whose occupancy (or not) distinguishes those charges.  

We can also contemplate incommensurate density waves.  This allows domain walls where the change in ${\rm Arg}$, and thus the accumulated charge, takes any value, rational or irrational. 

The possibility of zero-energy states at the termination of a 1-dimensional lattice was an early discovery of Shockley \cite{shockley}.  It gave a microscopically-based model of Bardeen's theory of ``surface states'', which in turn played an important role in the discovery of solid-state transistors.   Shockley's model is closely related to the models with which we began our discussion.  We can put his discovery into the same framework, and generalize it, by considering how we might model boundaries of chains (or, of course, surfaces) as extreme versions of domain walls, as follows.   

Taking Eqn.\,(\ref{electronLagrangian}) as our basic model, we can have some value $g\phi_+$ (not necessarily real) for the effective mass in bulk, in for $x \, >  \, 0$, while taking $g\phi_- \, \rightarrow \, \infty$ for $x \, < \, 0$, to make it difficult for the electrons to penetrate there.   (If desired we can have a small bridging region, and let $\phi$ interpolate continuously between those values.)   Thus, we realize the boundary as a specific kind of domain wall, to which our general analysis can be applied.   

Topological insulators \cite{TI}, in their simplest form, fit neatly into this framework, as follows.   Assuming $T$ symmetry, we let $g$ and $\phi$ be real.   Then the relevant issue is the relative sign of $m(x)$ for between $x<0$ and $x>0$.  If the sign changes, we have a zero energy surface state; if not, not.    If we keep spin as a passive (bookkeeping) quantum number throughout, then we will get even numbers of zero energy states, as in Eqn.\,(\ref{funnySpinCharge}); but if there are significant spin-dependent forces, then spin is not a good label, and in general we will find an odd number.   One simple version of how this can occur is to let the ``vacuum'' mass at $x \rightarrow - \infty$ be spin-dependent, asymptotically in the form $\sigma_3 m(x)$, with $m(x) \rightarrow \infty$ rapidly as $x \rightarrow - \infty$ and $m(x)$ approaches a constant -- possibly zero -- as  $x \rightarrow + \infty$. These constructions highlight that there need be no bulk signature of the topological insulator state.  Of course, we could write things alternatively so as to put the exotic structure in the bulk, which might (or might not) correspond to what emerges, in realistic cases, from a conventional band theory calculation.  

If we drop the requirement of $T$ symmetry, we bring in possibilities for surface fractional charge, according to the same basic mechanisms, exploiting the domain wall $\rightarrow$ boundary principle.  Note that the $T$ breaking can be a purely surface effect, but of course it need not be.  In the presence of an appropriate discrete symmetry the surface charge can be quantized, but in general it will be irrational.  In higher-dimensional situations we would predict, taking the models at face value, a surface charge density.   When we include Coulomb energy, there will be a strong incentive for the material to enforce neutrality on large scales, for example by fractionally populating conduction band states near the surface.   

\section{A Mechanical Monopole}

The ability of falling cats and of divers to re-orient themselves by changing shape, despite having zero angular momentum, is an instructive puzzle whose mathematical solution includes an elegant emergent gauge structure.  Let me briefly recall how this works \cite{sw}.   

In describing how much a deformable body has rotated, we immediately confront a basic issue, that no rigid rotation can connect different (non-congruent) shapes.  To make comparison possible, we can set up ``reference shapes'' with definite positions, and compare the position of our body, with its current shape, to the position of the associated reference shape.  This procedure introduces two issues:
\begin{itemize}  
\item It introduces an element of convention, because one is free to place the reference shapes differently.   If we write the relationship between body shapes and reference shapes as 
\begin{equation}\label{reference_shapes}
r^{(j)}(t) ~=~ {\cal R} (t) s^{(j)}(r^{(j)}(t))
\end{equation}
then a re-definition
\begin{equation}
{\tilde s}^{(j)} ~=~ {\cal U}(s^{(j)}) \, s^{(j)}
\end{equation}
induces the re-definition
\begin{equation}
\tilde{\cal R}(t) ~=~ {\cal R}(t) \, {\cal U}^{-1} (s(r(t)))
\end{equation}
where we have, mercifully, dropped the superscripts.  The net rotation between shapes at times $t_f, t_i$ is described alternatively as 
\begin{equation}
{\cal R}(t_f) \, {\cal R}^{-1}(t_i)
\end{equation} 
or 
\begin{equation}
\tilde {\cal R}(t_f) \, \tilde {\cal R}^{-1}(t_i) ~=~ {\cal R}(t_f) \, {\cal U}^{-1} (s(r(t_f)) \, {\cal U}(s(r(t_i)) \, {\cal R}^{-1} (t_i)
\end{equation} 
If the initial and final shapes are the same then the ${\cal U}$ factors cancels, and the ambiguity disappears.  
\item It fails when the shape degenerates to a point or a line.  In those cases there are congruences which leave the shape invariant, so the transformation to a reference shape is ambiguous.
\end{itemize}
The first point indicates that we have introduced a gauge structure, while the second indicates the possibility of singularities.

Having that background in mind, we are prepared to discuss the general relation between shape changes at zero angular momentum and re-orientation in space.   
With the transformation Eqn.\,(\ref{reference_shapes}) the zero angular momentum condition
\begin{equation}\label{zeroL}
0 ~=~ \sum\limits_j \, m^{(j)} r^{(j)} \times {\dot r}^{(j)}
\end{equation}
becomes 
\begin{equation}\label{shapeEquation}
0 ~=~ \sum\limits_j \, m^{(j)} s^{(j)} \times {\dot s}^{(j)} + \sum\limits_j \, m^{(j)} s^{(j)}\times  {\cal R}^{-1} \dot{\cal R} {s}^{(j)}
\end{equation}
Here the first term defines an effective angular momentum $L^{\rm shape}$, while ${\cal R}^{-1} \dot{\cal R}$ defines an effective angular velocity, in tensor form.   

With a bit of algebra we can cast Eqn.\,(\ref{shapeEquation}) into the form
\begin{equation}\label{connectionFormula}
V({\cal R}^{-1} \dot{\cal R}) ~=~ {\hat I}^{-1}\,  L^{\rm shape}
\end{equation}
where
\begin{eqnarray}
I_{pq} ~&=&~  \sum\limits_j \, m^{(j)} s^{(j)}_p {s}^{(j)}_q \\
\hat I_{pq} ~&\equiv&~ (I_{ll} \delta_{pq} - I_{pq} ) 
\end{eqnarray}
defines an effective inertia tensor and
\begin{equation}
V(M_{jk})_l ~\equiv~ \frac{1}{2} \epsilon_{jkl} M_{jk}
\end{equation}
defines the vector equivalent of the antisymmetric matrix $M$.  Equation (\ref{connectionFormula}) relates effective angular momentum and velocity through an effective inertia tensor.   The division between effective angular momentum and velocity are gauge (i.e. reference shape) dependent, but the effective inertia tensor is not.  

This construction provides a gauge connection in shape space, according to
\begin{equation}
{\cal R}^{-1} \dot{\cal R} \, dt  ~\equiv~ A_j \, dx^j
\end{equation}
The rotation resulting from a given trajectory in shape space is given by the ordered line integral (``Wilson line'') of this connection.  Note that $\frac {d}{dt}$ cancels, so that the accumulated rotation depends only on the geometry of the trajectory in shape space, not on how fast it is traversed.  Alternatively, we can say that it is time-reparameterization invariant.

Now let us turn to evaluating the connection for a system of three point masses.  For our standard shapes, we can put the first mass at the origin, the second at distance $\sqrt{m_2 r} \equiv \kappa \cos \theta$ along the $x$ axis, and the third in the $x-y$ plane at distance $\sqrt{m_3 s } \equiv \kappa \sin \theta$ and azimuthal angle $\phi$.  $\theta$ ranges between 0 and $\frac{\pi}{2}$.  Both $\theta$ and $\phi$ are ill-defined when $\kappa =0$, when all the masses coincide.  At that singular shape the preceding procedures to obtain $V$ fail, so we should only consider shape trajectories which avoid it.  When the distribution of masses is along a line, i.e. at $\phi = 0$, we have the ambiguity mentioned previously.  We can relieve it by demanding that the motion is restricted to a plane, which for simplicity let us do.  

Even then, since $\phi$ is ill-defined for $\sin \theta = 0$, we will need to use a different parameterization to cover that case.   A natural choice is to interchange the role of the second and third masses.  Calling the new parameters $\tilde \theta, \tilde \phi$ (with $\tilde \kappa = \kappa$), we have
\begin{eqnarray}
\tilde \theta &=& \frac{\pi}{2} - \theta \nonumber \\
\sin \tilde \theta &=& \cos \theta \nonumber \\
\tilde \phi &=& 2\pi - \phi
\end{eqnarray}
whenever both make sense.

Given these definitions, a simple calculation yields
\begin{equation}
L^{\rm shape} ~=~ \left(\begin{array}{c}0 \\0 \\  \kappa^2 \sin^2 \theta \, \dot \phi \end{array}\right)
\end{equation}
while $\hat I$ has the form
\begin{equation}
\hat I ~=~ \left(\begin{array}{ccc} * & * & 0 \\ {*} & * & 0 \\0 & 0 & \kappa^2 \end{array}\right)
\end{equation}
so that 
\begin{equation}
\hat I^{-1} ~=~ \left(\begin{array}{ccc} * & * & 0 \\ {*} & * & 0 \\0 & 0 & \kappa^{-2}\end{array}\right)
\end{equation}
and finally
\begin{equation}
V({\cal R}^{-1} \dot{\cal R}) ~=~ \left(\begin{array}{c} 0 \\ 0  \\ \sin^2 \theta \, \dot \phi\end{array}\right)
\end{equation}

(One might be concerned that our choice of shapes does not enforce that the center of mass stays fixed.  However, since under a change of origin the rigid motion $({\cal R}, \vec a)$ -- applying first the rotation $\cal R$ and then the translation $\vec a$ -- changes into $({\cal R}, {\cal R} \vec b - \vec b + \vec a)$, our determination of $\cal R$ is unaffected.  We could include the necessary compensating translations as a gauge potential, too, taking us into the group of rigid motions rather than just rotations.  But because the effect of enforcing the center of mass constraint vanishes for any closed path in shape space, the added piece introduces no curvature, and for simplicity I will discuss it no further here.)  

Since the motion is planar, we can regard interpret this result for the connection as an $SO(2)$ vector potential 
\begin{equation}
A_\phi (\rho, \theta, \phi) ~=~ \sin^2 \theta 
\end{equation}
on our shape space, with the accumulated rotation angle (relative to the standard shape) for a given trajectory in shape space given by
\begin{equation}
\Delta \alpha ~=~ \int A_\phi d\phi ~=~ \int\limits_{\rm initial}^{\rm final} \sin^2 \theta \, d\phi 
\end{equation}
integrated over the trajectory.  

The only non-vanishing component of the field strength associated with this potential is
\begin{equation}
F_{\theta \phi} ~=~ 2 \sin \theta \, \cos \theta
\end{equation}

The potential and field strength are independent of $\kappa$, but develop a singularity at $\kappa = 0$, where $\theta$ is ill defined.  The lack of $\kappa$ dependence reflects that we may re-scale (dilate) the shape at any time without affecting the overall rotation dynamics.  Recall that $\kappa = 0$ corresponds to the degenerate shape of three coincident points, where singularity might be anticipated.

Now let us fix $\kappa$ and evaluate the total flux emanating from the singularity.  We find
\begin{equation}\label{integratedFlux}
\oint F_{\theta \phi} \, d\theta \, d\phi ~=~ \int\limits_0^{\frac{\pi}{2}} 2 \sin \theta \, \cos \theta \, d\theta \int\limits^{2\pi}_0 d\phi ~=~ 2 \pi
\end{equation}
Here the limits on the $\theta$ integral are obvious from its definition, but the limits on the $\phi$ integral call for comment.  Indeed, one might be tempted to restrict the integral to $0 \leq \phi \leq \pi$, since the shapes with $\phi$ and $2\pi - \phi$ are congruent.  That congruence is, however, not implemented by an $SO(2)$ rotation, but requires reflection in the $x$ axis.  To work with continuous $SO(2)$ rotations, we must use a patching construction, which effectively distinguishes $\phi$ from $2\pi - \phi$.   (These reflective congruences can be implemented continuously within $SO(3)$, but not in a way that applies continuously to all three-dimensional realizations of the base shapes.  Note that while three-dimension inversion, as opposed to plane-dependent reflection, can be implemented globally, it does not make the relevant planar configurations equivalent.)  Equation (\ref{integratedFlux}) indicates that our connection has, for each fixed value of $\kappa$, the magnetic flux corresponding to the minimal Dirac charge.  Thus, we can identify the singularity at $\kappa = 0$ as a magnetic monopole in shape space.  

We can also bring the monopole structure out more topologically, by examining the patching construction required to knit together the $\tilde \theta, \tilde \phi$ and the $\theta, \phi$ coordinates.  We have 
\begin{eqnarray}
A_\phi &=& \sin^2 \theta \nonumber \\
A_{\tilde \phi} &=& - \sin^2 {\tilde \theta} = - \cos^2 \theta
\end{eqnarray}
The two potentials are related by a globally non-trivial gauge transformation, according to
\begin{equation}
A_\phi ~=~ \partial_\phi \,  \phi + A_{\tilde \phi}
\end{equation}

Finally, let us note that in terms of the alternative angle $\lambda \equiv \frac{\theta}{2}$ then we have conventional spherical co-ordinates $\lambda, \phi$ on our constant $\kappa$ spheres.  On those spheres, with the canonical metric, the magnetic flux is uniform.

\section{A Fractionated Molecule}

Upon considering the quantum version of this problem we encounter a related but different emergent gauge field, which illustrates another interesting phenomenon.

To keep things simple, in our system of three planar bodies let us take one to to be heavy and fixed at the origin, and the other two to be at fixed distances from the origin.  Then we are reduced to two dynamical degrees of freedom, i.e. two angles $\phi_1, \phi_2$.  Assigning them moments of inertia $I_1, I_2$, we have the rotational kinetic Hamiltonian
\begin{equation}\label{rotationalHamiltonian}
H_{\rm rot.} ~=~ \frac{p_1^2}{2I_1} + \frac{p_2^2}{2I_2} 
\end{equation}
with 
\begin{eqnarray}
p_1 ~&\equiv&~ -i\frac{\partial}{\partial \phi_1} \nonumber \\
p_2 ~&\equiv&~ -i\frac{\partial}{\partial \phi_2}
\end{eqnarray}

We can re-write Eqn.\,(\ref{rotationalHamiltonian}) as 
\begin{eqnarray}\label{separatedRotationalHamiltonian}
&\frac{1}{2(I_1 + I_2)}\, (p_1 + p_2)^2 \, +  \nonumber \\ 
& \, \frac{I_1 + I_2}{2I_1I_2} \bigl(\frac{p_1 + p_2}{2}\frac{I_2 - I_1}{I_1 + I_2} + \frac{p_1 - p_2}{2}\bigr)^2
\end{eqnarray}
Here $p_1 + p_2$ is the total rotation generator.  In our previous notation, it rotates the angle $\alpha$.  Now let us separate $\alpha$, as is appropriate for a rotationally invariant system, and impose the quantization condition
\begin{equation}
p_1 + p_2 ~=~ m
\end{equation}
Then in the second part of expression (\ref{separatedRotationalHamiltonian}) we have the bracketed term
\begin{equation}\label{shapeSpaceL}
(\frac{m}{2} \frac{I_2 - I_1}{I_1 + I_2} +  \frac{p_1 - p_2}{2}\bigr)^2 ~\rightarrow~ (p_2 - m\frac{I_2}{I_1 + I_2})^2 
\end{equation}
But $p_2$ generates changes in the relative angle in shape space, i.e., our previous $\phi$.  Thus, we see that non-trivial angular momentum for $\alpha$ generally yields fractional (kinetic) angular momentum \cite{fw} for the shape-space angle $\phi$.  The fractional part is
\begin{equation}\label{fractionalPart}
\delta l ~=~ \frac{mI_2}{I_1 + I_2}
\end{equation}
Note that this approaches an integer for $I_1/I_2 \rightarrow 0$ or $\rightarrow \infty$, as it should.

The bracketed term in Eqn.\,(\ref{separatedRotationalHamiltonian}) might look strange at first sight, but if we substitute $p_1 = I_1\omega_1, p_2 = I_2\omega_2$ we find that it is proportional to $(\omega_2 - \omega_1)^2$.  Thus, it represents the contribution of relative angular velocity.   It is only the passage from Lagrangian to Hamiltonian that introduces complications.  But of course quantum theory requires the Hamiltonian.

In an alternate description, corresponding to the other patch in our preceding monopole construction, we interchange the roles of bodies 1 and 2.  In that way, we replace Eqn.\,(\ref{shapeSpaceL}) by 
\begin{equation}
 (\frac{m}{2} \frac{I_2 - I_1}{I_1 + I_2} +  \frac{p_1 - p_2}{2}\bigr)^2 ~\rightarrow~ (p_1 + m\frac{I_1}{I_1 + I_2})^2 
\end{equation}
and hence Eqn.\,(\ref{fractionalPart}) by
\begin{equation}\label{newFractionalPart}
\delta l ~=~ - \frac{mI_I}{I_1 + I_2}
\end{equation}
Consistency requires that the difference between Eqn.\,(\ref{fractionalPart}) and Eqn.\,(\ref{newFractionalPart}) must be an integer, and indeed it is the integer $m$.  
Thus we see an interesting connection between the monopole in shape space, which we originally uncovered classically, and the quantum theory of the same mechanical system.

Another consistency check is to consider the free limit, where Eqn.\,(\ref{rotationalHamiltonian}) is the entire Hamiltonian.  Then of course by substituting integer values for $p_1, p_2$ we get the spectrum, either directly from Eqn.\,(\ref{rotationalHamiltonian}) or after some algebra from Eqn.\,(\ref{separatedRotationalHamiltonian}), with identical results.   The significance of the separation procedure is that overall rotational symmetry is more generic than shape independence, and by exploiting it we always reduce the dimensionality of the potentially non-trivial dynamics.        

If we restore the dynamics that allows $I_1, I_2$ to vary, then the preceding construction gives an emergent dynamical compact $U(1)$ gauge field in shape space, which however is not governed by the classic Maxwell-Yang-Mills action.  In that interpretation, the total angular momentum supplies an effective charge.

\bigskip
\bigskip

{\it Acknowledgement}: The first section partly adapts material from \cite{fw_nobel}.   The second section adapts material from a forthcoming paper with X. Peng, J. Dai, and A. Niemi \cite{pdnw}.  This work is supported by the U.S. Department of Energy under grant Contract  Number DE-SC0012567, by the European 
Research Council under grant 742104, and by the Swedish Research Council under Contract No. 335-2014-7424.

\end{document}